\documentclass[fleqn,usenatbib]{mnras}

\usepackage{newtxtext,newtxmath}

\usepackage[T1]{fontenc}

\DeclareRobustCommand{\VAN}[3]{#2}
\let\VANthebibliography\thebibliography
\def\thebibliography{\DeclareRobustCommand{\VAN}[3]{##3}\VANthebibliography}

\usepackage{graphicx}	
\usepackage{amsmath}	
 \usepackage[usenames,dvipsnames]{xcolor}

\usepackage[normalem]{ulem}

\title[Hot Subdwarf Triples]{Forming Hot Subluminous Stars from Hierarchical Triples - I. The Role of an Outer Tertiary on Formation Channels}

\author[Holly P. Preece]{
Holly P. Preece,$^{1}$\thanks{E-mail: hpreece@mpa-garching.mpg.de}
Adrian S. Hamers,$^{1}$
Tiara Battich$^{1}$,
Abinaya Swaruba Rajamuthukumar$^{1}$
\\
$^{1}$Max Planck Institute for Astrophysics, Garching
}

\date{Accepted XXX. Received YYY; in original form ZZZ}

\pubyear{2022}

\begin{document}
\label{firstpage}
\pagerange{\pageref{firstpage}--\pageref{lastpage}}
\maketitle

\begin{abstract}
 We present evolutionary pathways for creating hot subdwarf OB (sdOB) stars from hierarchical triple configurations. We use the population synthesis code Multiple Stellar Evolution (MSE) to follow the stellar, binary, and gravitational dynamical evolution of triple-star systems. To ascertain the effect of the outer tertiary, we also consider the evolution of the inner binary with the tertiary component removed. We find we are able to create sdOB stars in single, binary and triple configurations. We also demonstrate that it is possible to form sdOBs in systems which undergo triple common envelope evolution, when the tertiary star undergoes unstable mass transfer onto the inner binary. We are unable to create single or wide sdOB systems without involving a merger earlier in the evolution. The triples can produce sdOBs in binaries with wide, non-interacting companions through binary interactions, which is impossible in isolated binaries. Owing to the closeness of the inner binary in hierarchical triples the formation channels associated with stable mass transfer are suppressed when compared to the isolated binary case. 
\end{abstract}




\section{Introduction}

Hot subluminous O (sdO) and B (sdB) type stars are evolved, low-mass stars which sit on the extreme horizontal branch (EHB). The sdBs are observed to have surface gravities $5 < \log_{10}g_{\rm{surf}} < 6$ and surface temperatures $20000 < T_{\rm{eff}}/\rm{K} < 40000$ \citep{discoveryzwicky, specdefinition, heberehb}. In addition, they are chemically peculiar. The stars are both observed and predicted to have masses peaked around $0.47 \, M_\odot$ \citep{2022arXiv220702001S}. Most of the sdOBs are thought to be core Helium (He) burning stars with low mass Hydrogen (H) envelopes. The radially extended H rich envelopes have masses < $0.01 \, M_\odot$ \citep{2016PASP..128h2001H}. Extreme mass-loss or binary interaction is required to remove the envelope of the star. 

The single star evolution channels include enhanced wind on the red giant giant branch \citep{1996ApJ...466..359D}, or high initial He composition ($\sim 0.4$ by mass fraction, \citealt{2017A&A...597A..67A}). A number of binary evolutionary channels have been proposed to form an sdOB \citep{hani,hanii,podsiformation08,hewdsimon,hewdmszhang,wdmsmergeclausen}. Stable or unstable Roche lobe overflow (RLOF) at the tip of the red giant branch (RGB) or the merging of two stars, typically white dwarfs (WDs), are the accepted binary channels. The sdOB progenitor stars have masses between $1\,M_\odot$ and $4\,M_\odot$. In many cases He is ignited degenerately via the He-flash. About half of the sdOBs are observed to be in binaries. The majority of these binaries are in close orbits and have orbital periods orbital periods less than ten days \citep{maxtedbinaries,napiow04binaryf,binaryfraction2011,2022arXiv220702001S}. 

The close sdOB binaries originate from common-envelope evolution \citep{hani, hanii}. The companions to the sdOBs are either low-mass main-sequence stars (dM) or WDs \citep{2022arXiv220702001S}. If the sdOB progenitor is initially the most massive star an sdOB+dM system is formed. If the sdOB progenitor is initially less massive than its companion an sdOB+WD binary forms. For the sdOB+dM channel, the more massive star of the binary evolves onto the red giant branch. Common-envelope evolution occurs after unstable RLOF. He is ignited in the core to form an sdOB. To form an sdOB+WD the initially more massive primary evolves to become a white dwarf. Stable or unstable RLOF typically occurs as the primary evolves into a WD. After the primary has evolved onto the WD branch the secondary evolves onto the RGB. Unstable mass-transfer leading to common-envelope evolution is triggered. The envelope of the RGB star is removed. He is ignited in the core of the stripped star to form an sdOB binary with a WD companion.

The wide sdOB binaries form through stable Roche Lobe overflow \citep{vosi, vosii, vosiii}. The companions are typically F/G/K main-sequence stars. The star which forms the sdOB is initially the most massive star in the binary. It evolves off the main-sequence and ascends the RGB. The star fills its Roche Lobe and stable mass-transfer begins. The H envelope is fully removed during this episode of mass-transfer, then He is ignited in the core. 

Mergers are the third binary avenue predicted to produce sdOBs. In this case, the sdOB produced is a singleton. Numerous merger channels have been suggested. \cite{webbinkhewd,hewdmszhang} both proposed the merging of two He; the remnant then reignites He to form an sdOB. \cite{justhamco} showed that a hybrid CO-He WD and He WD can merge wherein the more massive hybrid WD accretes the He WD. He burning resumes in a shell around the CO core of the merged product. If the He WD is more massive than the hybrid CO-He WD then the He WD can accrete the companion to create a CO rich subdwarf with a He burning core \citep{marcellowd}. \cite{wdmsmergeclausen} show that a sdOB can result from a He-WD merging with a low-mass main-sequence star. \cite{sokermerger, politanocemerger} posit that the post-common envelope merging of a red giant core with a low-mass main-sequence can give rise to a single sdOB.

Several sdOBs in hierarchical triples candidates have been observed. Typically, the sdOB is part of the inner binary, with an unseen companion. The first candidate hierarchical triple system containing an sdOB was PG 1253+284 \citep{hebertriple}. SDSS J095101.28+034757.0 exhibits IR excess from a wide companion and was the second proposed hierarchical triple sdOB system \citep{kupfertriple}. \cite{pelisoli2020} identified 6 proper motion pairs with evidence for an additional close companion to the sdOB in the system. 

For solar type stars approximately 10\% of systems are observed to be hierarchical triples and 30\% of systems are expected to be binaries \cite{moepq}. The relative fraction of triple systems to binary systems increases with mass For $\gtrsim 20 \,\mathrm{M}_\odot$ primaries, the relative fraction is $\sim 1:1$. At $5\,M_\odot$ $\sim$24\% of systems are triples and $\sim$36\% are binaries. One would expect between $\sim$25\% and $\sim$40\% for sdOB progenitor systems to have been hierarchical triples, yet this area remains relatively unstudied. 

Hierarchical triples consist of a close inner binary with a wide tertiary component. In stable configurations, orbital averaged secular approximations derived from Hamiltonian mechanics can be employed. The inner binary is subject to the usual binary interactions however the outer tertiary, even if in a very wide orbit, can have a substantial influence on the orbital evolution. If the outer tertiary is sufficiently inclined relative to the inner binary, von Zeipel-Lidov-Kozai (ZLK) oscillations can be triggered \citep{vonzo,lidovo,kozaio,noazzlk}. During ZLK oscillations, orbital angular momentum is exchanged between the two orbits due to secular torques. The eccentricity of the inner orbit is excited and the relative inclination between the binary and tertiary varies. When the eccentricity is excited the periastron distance decreases. In combination with tidal interactions, ZLK oscillations are efficient at shrinking orbits \citep{mazehzlk,eggletonzlki,eggletonzlkii,fabrickyzlk}. The ZLK oscillations can also enhance merger rates \citep[e.g.,][]{2002ApJ...578..775B, 2011ApJ...741...82T,2017ApJ...841...77A, 2018ApJ...863...68L, 2018ApJ...864..134R, 2018A&A...610A..22T} and induce mass-transfer \citep[e.g.,][]{2019ApJ...882...24H,2020A&A...640A..16T}. Mass-loss and mass-transfer between stellar components can alter the orbits and trigger phases of dynamical instability \citep{kiseleva3star, iben3star,portegeis3star, perets3star,tedi,2022A&A...661A..61T}. If the system becomes dynamically unstable the gravitational dynamics are complex and chaotic. Exchanges, ejections, collisions and mergers can all occur.

In this paper, we examine the formation channels for sdOB stars from hierarchical triples using population synthesis methods. We aim to identify the  evolution channels which lead to an sdOB star and the configurations of their host systems. The focus of this paper is to find possible evolutionary pathways. Follow up work shall focus more on the statistical properties of the found evolutionary pathways. In Section 2 we discuss the stellar models used in this work. Details of the population synthesis code and the initial conditions used are included. In Section 3 we present new formation channels discovered in this work. We examine how the systems are formed and whether the sdOB is observed to be in a triple, binary or is a singleton. Section 4 presents the discussion and Section 5 concludes.

\section{Population Synthesis}
The code used for the simulations in this paper is Multiple Stellar Evolution (MSE), a population synthesis algorithm to model the dynamical, stellar, and binary evolution of multiple-star systems with any multiplicity\footnote{The version used in this paper is v0.87.}. MSE is written in C/C++ and has a convenient Python interface; it is publicly available\footnote{\href{https://github.com/hamers/mse}{https://github.com/hamers/mse}}. A succinct description of the main features of the code is given here; for details, we refer to \citet{2021MNRAS.502.4479H}. 

The gravitational dynamical evolution of the multiple system is modelled using either a secular (i.e., perturbative and orbit-averaged) approach \citep{2016MNRAS.459.2827H,2018MNRAS.476.4139H,2020MNRAS.494.5492H} when the latter approximation applies, or via highly accurate direct $N$-body integration using the algorithmic chain regularisation code MSTAR \citep{2020MNRAS.492.4131R}. The code dynamically switches between both methods depending on the current state of the system. In both cases, post-Newtonian (PN) correction terms are taken into account up to and including 2.5 PN order. 

The evolution of single stars in MSE is taken into account by adopting the fast analytic fitting formulae to 1D stellar evolution models developed by \citet{2000MNRAS.315..543H}. Included are mass loss due to stellar winds, and spin-down due to magnetic braking. Massive stars evolving to become NSs or BHs are assumed to receive a natal kick velocity, with the assumed kick distributions given by Model 1 of \citet{2021MNRAS.502.4479H}, i.e., a Maxwellian distribution with $\sigma = 265 \, \mathrm{km\,s^{-1}}$ for NSs \citep{2005MNRAS.360..974H}, and $\sigma = 50 \, \mathrm{km\,s^{-1}}$ for BHs. Mass loss from stellar winds is assumed to act adiabatically on the orbits in the multiple system, whereas mass loss from supernova explosions is assumed to be instantaneous.

Tidal interactions between stars are taken into account by assuming the equilibrium tide model, with the tidal dissipation strength prescribed by \citet{2002MNRAS.329..897H}. Some details of mass transfer between two stars are modelled similarly to \citet{2002MNRAS.329..897H}, and include the mass transfer rate, aging/rejuvenation, and the conditions for mass transfer stability. The orbital response to mass transfer is modelled differently, however, by adopting the analytic model for eccentric mass transfer of \citet{2019ApJ...872..119H}. Unstable mass transfer is assumed to lead to CE evolution, which is modelled via the $\alpha$ CE formalism similarly to \citet{2002MNRAS.329..897H}. 

In tight multiple systems, an outer companion can transfer mass to an inner subsystem consisting of two or more stars. This process is modelled in MSE in an approximate way by adopting prescriptions and simplified simulations for both stable and unstable mass transfer. Both prescriptions are motivated by more detailed simulations \citep{2014MNRAS.438.1909D,2020MNRAS.498.2957C,2021MNRAS.500.1921G}. 

MSE also takes into account perturbations from fast-moving and distant field stars passing by the multiple system. These impulsive encounters are taken into account using a Monte Carlo approach. Typically, only wide orbits with semi-major axes in excess of $10^3\, \rm{AU}$ are significantly affected by these fly-bys. 

We use the supernova Ia prescription of \cite{2016A&A...589A..43N}, Rajamuthukumar et al. in prep. The prescription describes sub-Chandrasekhar detonation of a CO-WD after accreting matter from a He-WD companion, as well as supernova Ia explosions resulting from WD collisions.

\subsection{Initial Conditions}
We generate two data sets of 100,000 systems. The first data set are hierarchical triples (referred to as the triple case from here onward). The second data set are the inner binaries of the hierarchical triples with the outer tertiary removed (referred to as the binary case from here onward). The same initial conditions are used for the inner binaries for the binary and triple case. We compare the results of the triple case to the binary case to assess the impact of the outer tertiary on formation. 

\begin{figure}
 \includegraphics[width=0.98\columnwidth]{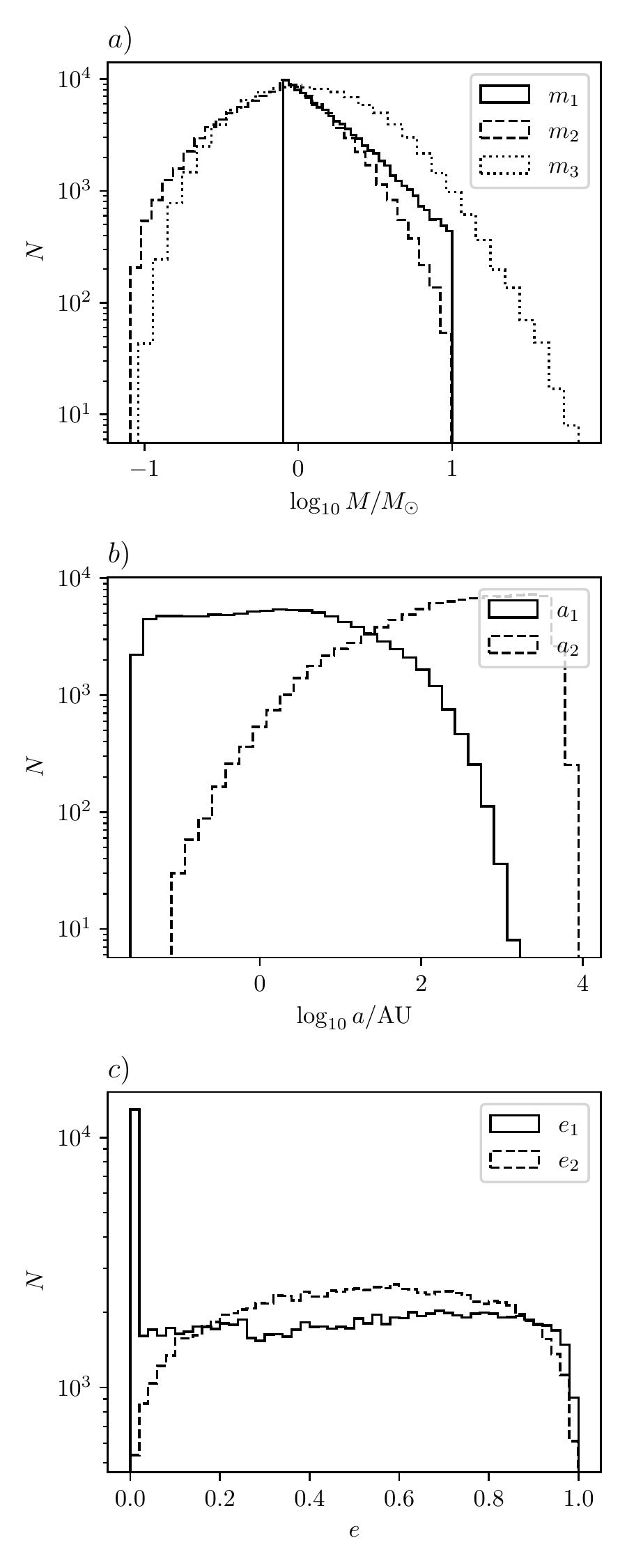}
 \caption{The initial conditions of the systems used for the calculations of this paper. Subplot $a)$ shows the masses of the stars where $m_1$ is the most massive star in the inner binary, $m_2$ is the least massive star in the inner binary and $m_3$ is the mass of the tertiary. Subplot $b)$ are the semi-major axes of the inner binary, $a_1$, and outer tertiary orbit, $a_2$. Subplot $c)$ are the eccentricities of the inner binary, $e_1$, and the outer tertiary, $e_2$. The same distributions of $m_1$, $m_2$, $a_1$ and $e_1$ are used for the data set with the outer tertiary removed.}
 \label{fig:ics}
\end{figure}

The distributions of initial parameters are shown in Fig \ref{fig:ics}. For this work, $m_1$ is the mass of the initially more massive primary star in the inner binary, $m_2$ is the mass of the initially less massive secondary star in the inner binary, $m_3$ is the mass of the outer tertiary, $a_1$ is the semi-major axis of the inner binary, $a_2$ is the semi-major axis of the outer tertiary, $e_1$ is the eccentricity of the inner binary and $e_2$ is the eccentricity of the outer tertiary. We limit the mass of the most massive star in the inner binary to the range $0.8 < M_1 / M_\odot< 10$. The secondaries have masses larger than $0.08\,M_\odot$. The tertiaries have masses in the range $0.08 < m_3/M_\odot < 100$. The orbital period of the inner binaries are in the range $0.2 < \,\log_{10} P_{\rm{orb,1}}/\rm{days} < 7.9$. Eccentricities of both orbits are between 0 and 1. 

The initial conditions are selected from the observational distributions obtained by \cite{moepq} using a Monte Carlo sampling method. The mass of the primary of the inner binary follows \cite{kroupa}. The mutual inclinations are uniform in $\cos \, i$. The longitude of the ascending node (LAN) and argument of periapsis are both uniform. We further allow for the outer tertiary to be more massive than the combined mass of the inner binary. Comparison to observational data shows these high mass outer tertiaries are approximately distributed according to a decreasing exponential \citep{tokovininq2big}. The Multiple Star Catalogue of \citet{tokovininq2big}, and by extension the distribution of $q_\mathrm{out} \equiv m_3/(m_1+m_2)$ assumed here, likely has major observational biases. We will consider the dependence on the assumed distribution of $q_\mathrm{out}$ in future work. 

The distribution of the outer tertiaries with mass ratios greater than 1 are described by a decaying exponential function which is of the form 
\begin{equation}
    \frac{\mathrm{d}N}{\mathrm{d}q_\mathrm{out}} \propto \exp(-q_\mathrm{out}\lambda)
\end{equation}
where $q_{\rm{out}}$ is the mass ratio of the outer tertiary to the inner binary and $\lambda$ = 1.05.
Owing to the dynamical stability criteria of \citet{2001MNRAS.321..398M}, the inner binaries of hierarchical triples, and thus of the binary case considered here, are in tighter orbits than true isolated binaries \citep[e.g.,][]{2019ApJ...883...23H}. As a consequence of these tight inner orbits, many systems achieve tidal circularization during the pre-main sequence evolution and so there is a spike of eccentricities at 0 \citep{moeprems}. A similar peak at $e=0$ is seen in the distribution of isolated binaries but is less pronounced. We restrict the initial mass of the more massive star in the inner binary to the range $0.8 < M_1/\,M_{\rm{\odot}}<10$. Comparison with results obtained from binary population synthesis show that the majority of sdOB stars are formed from primaries in this mass range. Owing to the large parameter space involved with simulating hierarchical triples, and the relative computational expense especially for tight triple systems, we restrict the initial conditions to those most likely to form sdOB stars. This is an exploratory study with a focus on finding evolutionary pathways which lead to sdOBs from initially hierarchical channels. Statistical properties for individual channels and the dependence on different code parameters shall be saved for future study.

To generate the hierarchical triples first the inner binary is sampled following \cite{moepq}. Next, the inner binary is treated as a point source and the the tertiary orbit is drawn from the same distribution. The stability criterion of \cite{2001MNRAS.321..398M} is applied to the triple to ensure it is in a dynamically stable hierarchical configuration. Any dynamically unstable systems are rejected. Systems that are initially Roche lobe overflowing (at periapsis) are also rejected.

\section{Formation Channels}
\begin{figure}
 \includegraphics[width=\columnwidth]{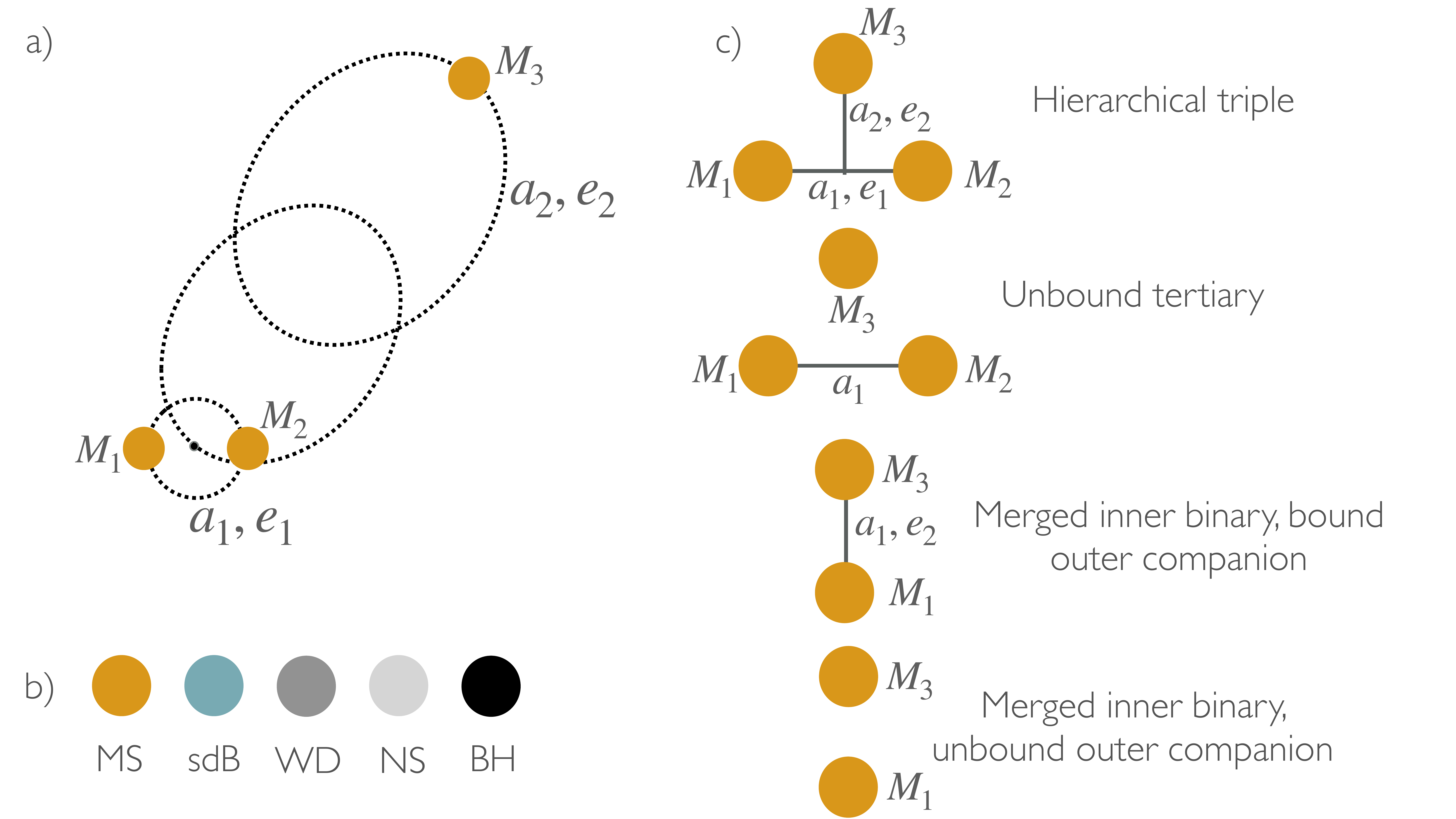}
 \caption{A schematic showing the diagrams style used to represent evolutionary channels. a) is a hierarchical triple for reference, $m_1$ is the primary and most massive star in the inner binary, $m_2$ is the secondary and least massive star in the inner binary, $m_3$ is the outer tertiary. The semi-major axis of the inner binary is $a_1$ and the outer tertiary is $a_2$. b) shows the colours used to represent different evolutionary stages. c) are various orbital configurations. If the inner binary, $m_2$ now represents the object which was originally the outer tertiary, if bound the semi-major axis is $a_1$.}
 \label{fig:refdiagram}
\end{figure}

Based on the outcome of the MSE simulations we present 14 distinct evolutionary scenarios for producing an sdOB star from a hierarchical triple with 31 sub-channels. Fig. \ref{fig:refdiagram} is a schematic outlining the diagrams used to represent the evolutionary outcomes. We group our results by the configuration of the system at the time of sdOB formation. We present hierarchical triples, close binaries, wide binaries and singletons separately. As in the binary evolution case, an sdOB can be made either by removing the envelope of an RGB star or merging two He-WDs. MSE does not currently have prescriptions to form sdOBs from mergers of He-WD + CO-WD or He-WD + low-mass MS stars. 

Given the uncertain physics of common-envelope evolution and mergers, and the more approximate nature of population synthesis calculations relative to detailed stellar models, we employ a broad criterion for an sdOB. In this work, sdOBs are defined as stripped core He-burning stars with masses below $0.65\,M_\odot$. We further require that the sdOBs survive for a minimum of $5\,\rm{Myrs}$ so as not to consider systems which merge with a close companion shortly after formation.

\subsection{Hierarchical Triple Channels}
\begin{figure}
 \includegraphics[width=\columnwidth]{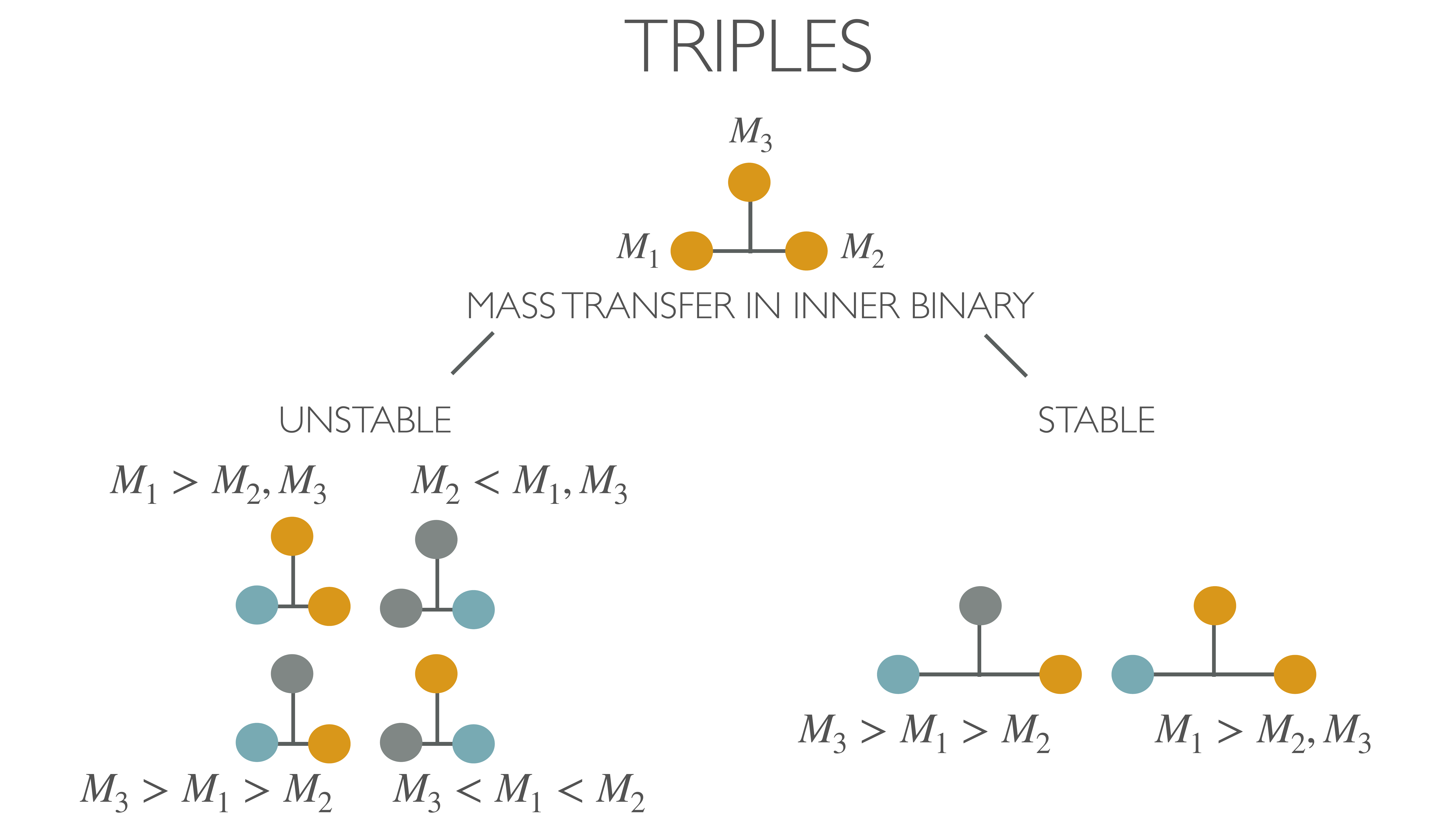}
 \caption{The formation channels identified for hierarchical triples.}
 \label{fig:sdbtriplesform}
\end{figure}

The sdOB stars in hierarchical triple configurations on the zero-age EHB (ZAEHB) are all the result of the removal of the envelope of an RGB star. The Roche lobe overflow can cause either stable or unstable mass-transfer. If the mass-transfer is unstable a common-envelope forms and expels the envelope of the RGB star. If the mass-transfer is stable the RGB envelope is transferred onto a main-sequence companion. An overview of the formation channels leading to hierarchical triple sdOBs is shown in Fig. \ref{fig:sdbtriplesform}. The formation channels are similar to the binary channels described in the introduction.

\subsubsection{Common Envelope}
We identify four possible combinations of companions to the sdOB stars formed via unstable Roche Lobe overflow. The combinations of primary, secondary and tertiary are sdOB + MS + MS, sdOB + MS + WD, WD + sdOB + MS, WD + sdOB + WD. In all cases one of the stars initially in the inner binary becomes an sdOB. The outcome depends on the relative initial masses of the three objects. If the outer tertiary is less massive than the star from the inner binary which forms the sdOB, it is still a main-sequence star when the sdOB is formed. If the outer tertiary is initially more massive than the sdOB forming star, it evolves into a WD before sdOB formation. 

If the initially more massive star of the inner binary is the object which becomes an sdOB the companion is a main-sequence star. The primary of the binary evolves more quickly than its lower mass companion. When it arrives on the RGB it fills its Roche lobe and mass-transfer commences. The mass-transfer becomes unstable and common envelope evolution ensues. The envelope is ejected and He is ignited in the core of the stripped star to make an sdOB. If the initially less massive star of the inner binary forms an sdOB the other component of the binary is a WD. The more massive star evolves into a white dwarf. Common-envelope occurs during one of the giant branches. The initially less massive star of the inner binary then evolves onto the red giant branch. On the RGB it undergoes a second phase common-envelope evolution with its WD companion. After the envelope is expelled the star ignites He and settles onto the extreme horizontal branch.

\subsubsection{Stable Roche Lobe Overflow}
We identify two possible companion configurations for sdOBs which form via stable Roche lobe overflow. The combinations of primary, secondary and tertiary are sdOB + MS + MS and sdOB + MS + WD. As with the common envelope scenario presented in the previous subsection, the outer tertiary can be either a MS or WD. 

If the tertiary is initially the most massive object in the system it is a WD at the time of sdOB formation. If the tertiary is less massive than the primary of the inner binary it is still a MS star at the time of sdOB formation. The star which forms the sdOB is always the more massive of the two stars in the inner binary. The primary of the inner binary evolves onto the RGB and overflows its Roche Lobe. Mass is transferred stably onto the main-sequence companion until the H rich envelope is removed. He is ignited in the core of the stripped star to form an sdOB. The semi-major axis of the inner binaries are wider than in the common-envelope case. 

\subsection{Close Binary Channels}
\begin{figure}
 \includegraphics[width=\columnwidth]{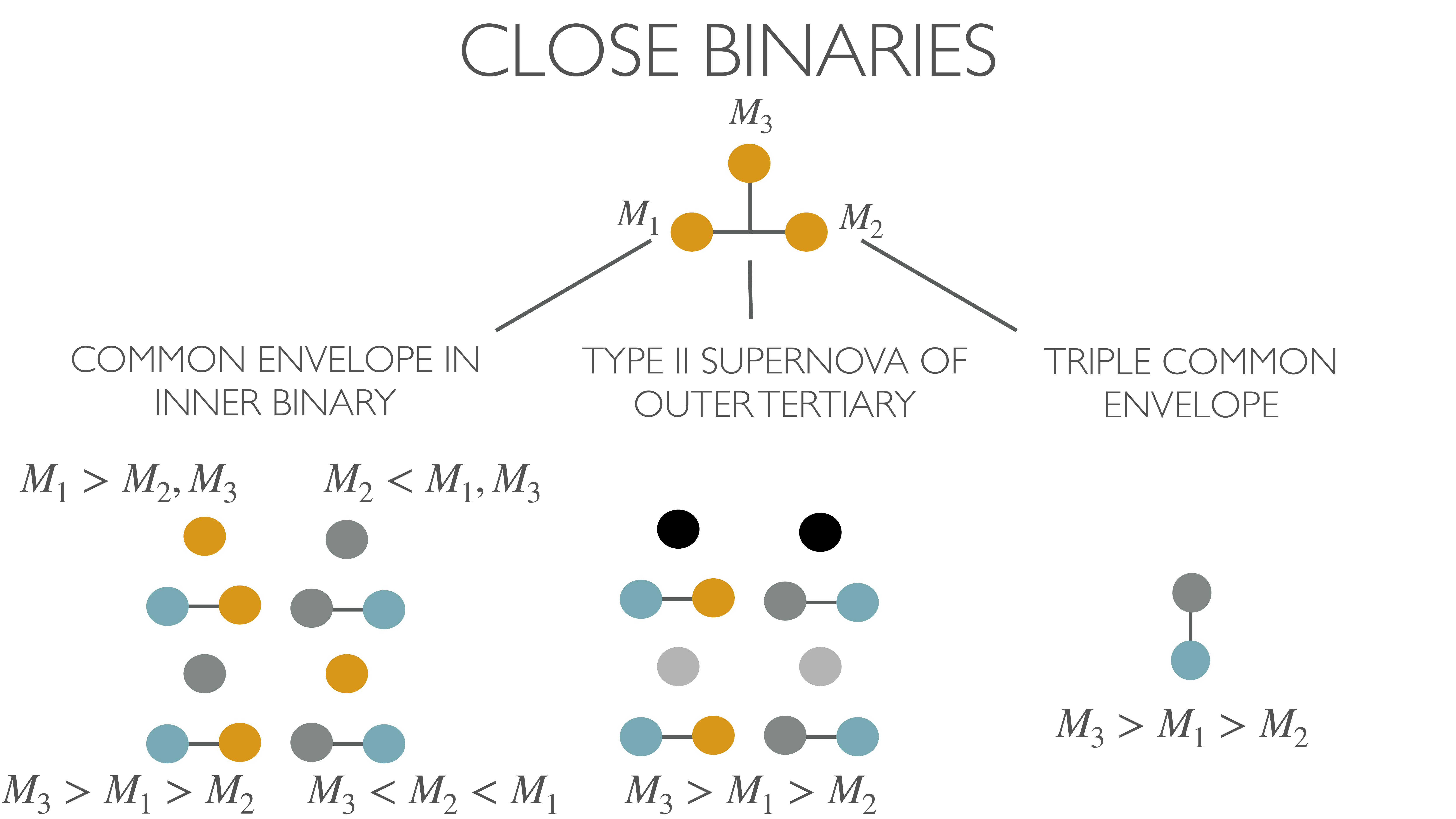}
 \caption{The formation channels identified for close binaries.}
 \label{fig:sdbbinariesform}
\end{figure}

The sdOBs in close binaries are formed via four distinct evolutionary pathways. An overview of the evolutionary channels to produce an sdOB star in a close binary is show in Fig. \ref{fig:sdbbinariesform}. Either the outer tertiary becomes unbound or the stars in the inner binary merge. The tertiary can be unbound owing to orbital instability during common-envelope evolution or if there is a Blaauw and/or natal kick from a type II supernova. The inner binary can merge during the early main-sequence evolution if the triple is initially very tight. The inner binary can also merge following a phase of triple common-envelope evolution.   

\subsubsection{Common-Envelope + Ejection}
These systems follow a similar evolution as the common-envelope channel to form the sdOB in a hierarchical triple. When the common-envelope is ejected a substantial amount of mass is removed from the inner binary. Consequently, the relatively wide outer tertiary is expelled. The sdOB star has either a main-sequence or WD companion bound in a tight orbit. 

If the primary of the inner binary forms the sdOB the companion is a MS star. The primary evolves onto the RGB, overflows its Roche lobe and undergoes common envelope evolution. He is ignited in the core and the primary becomes an sdOB star with a MS companion. 

If the secondary becomes the sdOB the companion is a WD. First, the primary of the inner binary evolves onto the WD branch. Some stable mass-transfer occurs on the giant branches. The secondary then evolves onto the RGB, undergoes common-envelope evolution. He is ignited in the core of the stripped star to create an sdOB + WD close binary. 

The unbound tertiary can be either a MS star, if it is initially less massive than the sdOB forming star, or a WD, if it is initially more massive than the sdOB forming star.

\subsubsection{Type II Supernovae of the Outer Tertiary}
If the outer tertiary is sufficiently massive, it evolves on a very short time-scale and undergoes a type II supernova. The mass loss and/or natal kick from the supernova unbinds the formed neutron star (NS) or black hole (BH) from the MS-MS inner binary. 

An sdOB is formed in the binary by removing the envelope of an RGB star via either common-envelope evolution or stable Roche lobe overflow. If the envelope is removed by stable Roche lobe overflow the sdOB has a MS companion. If the envelope is removed by common-envelope evolution the sdOB can have either a MS or WD companion.

\subsubsection{Triple Common Envelope + Common Envelope}
This formation channel occurs when the most massive star is the outer tertiary component and the triple is in an initially tight configuration. The outer tertiary evolves into a thermally pulsing asymptotic giant branch object. During the thermally pulsing phase it overflows its Roche lobe and mass-transfer begins. If the mass-transfer becomes unstable a common-envelope forms. The common-envelope engulfs both stars of the inner binary. The envelope is expelled and the tertiary star cools to form a CO-WD. The inner binary merges to form a more massive MS star. As the triple common-envelope shrunk the outer orbit substantially a CO-WD + MS close binary remains. The MS star later evolves onto the RGB, undergoes common-envelope ejection and an sdOB + CO-WD in a tight orbit is formed. 

After the sdOB star has exhausted its core He and becomes a CO-WD, the two CO WDs merge on timescales shorter than the Hubble time. The combined mass of the two CO WDs is typically > $1.2 \, M_{\rm{\odot}}$ thus these types of systems are promising Ia supernova progenitors. 

\subsubsection{Triple Common Envelope}
An sdOB can be formed directly from triple common-envelope. These systems tend to be in even tighter configurations than the triple common-envelope + common envelope channel. The outer tertiary, which is again the most massive star in the system, evolves onto the RGB. The RGB star overflows its Roche lobe and a common-envelope forms which engulfs the inner binary. The common envelope is ejected. The inner binary merges to form a more massive MS star. He is ignited in the core of the RGB stripped star. The result is an sdOB + MS close binary.

\subsubsection{Early Main-Sequence Merger}
If the hierarchical triple is initially so tight that it only just satisfies the stability criteria of \cite{2001MNRAS.321..398M}, the inner binary may merge during the early MS evolution. The wind mass-loss on the MS triggers a dynamical instability in the orbits which leads to a collision in the inner binary. Another possibility is that $e_2$ slightly increases due to secular evolution and triggers the dynamical instability. The former hierarchical triple is now a close binary with two MS stars. An sdOB can be formed via either stable or unstable RLOF in either object. The companion to the sdOB is either a MS or WD star.

\subsection{Wide Binary \& Singleton Channels}

\begin{figure}
 \includegraphics[width=\columnwidth]{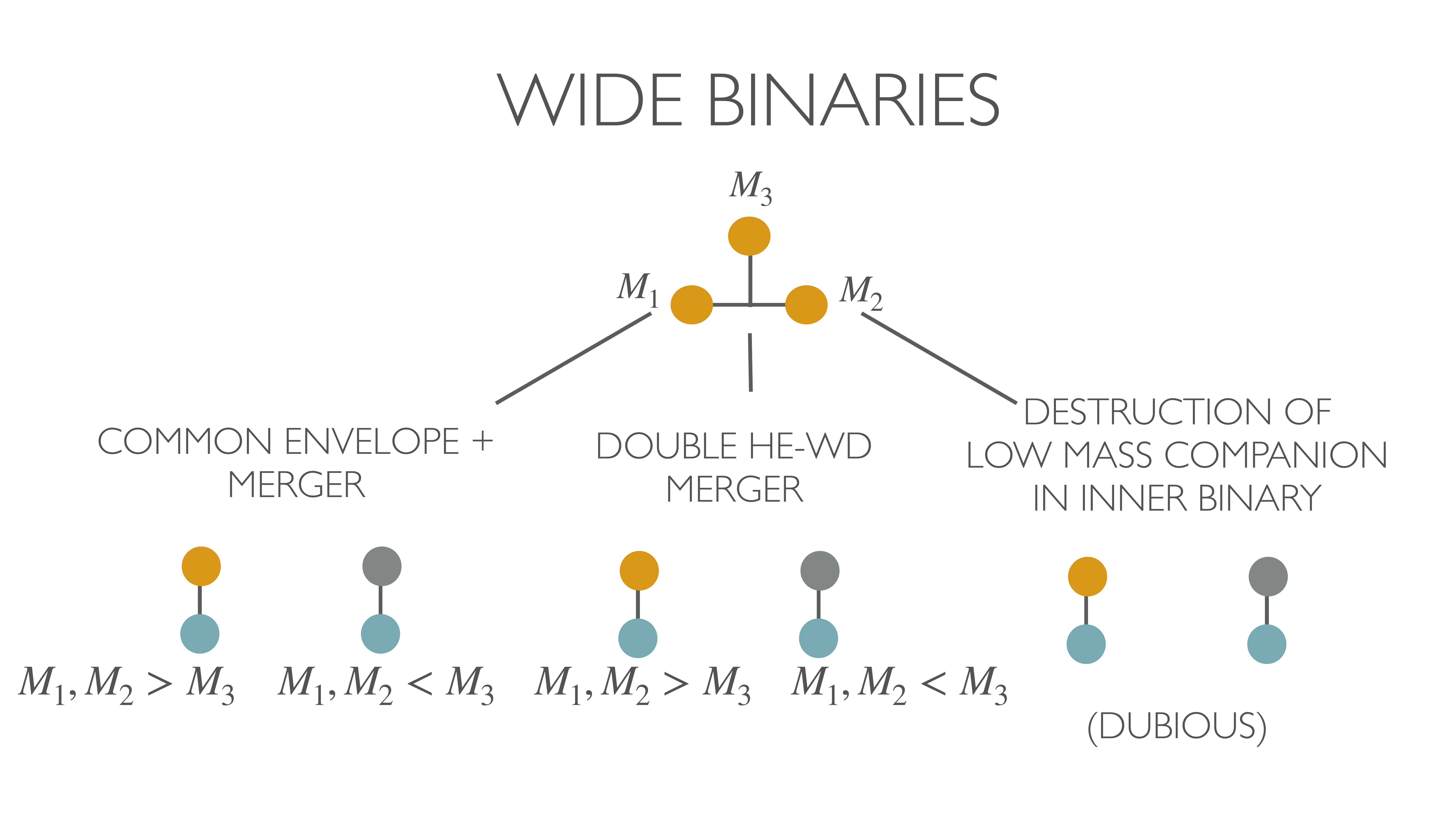}
 \caption{The formation channels identified for wide binaries.}
 \label{fig:sdbwidebinariesform}
\end{figure}

\begin{figure}
 \includegraphics[width=\columnwidth]{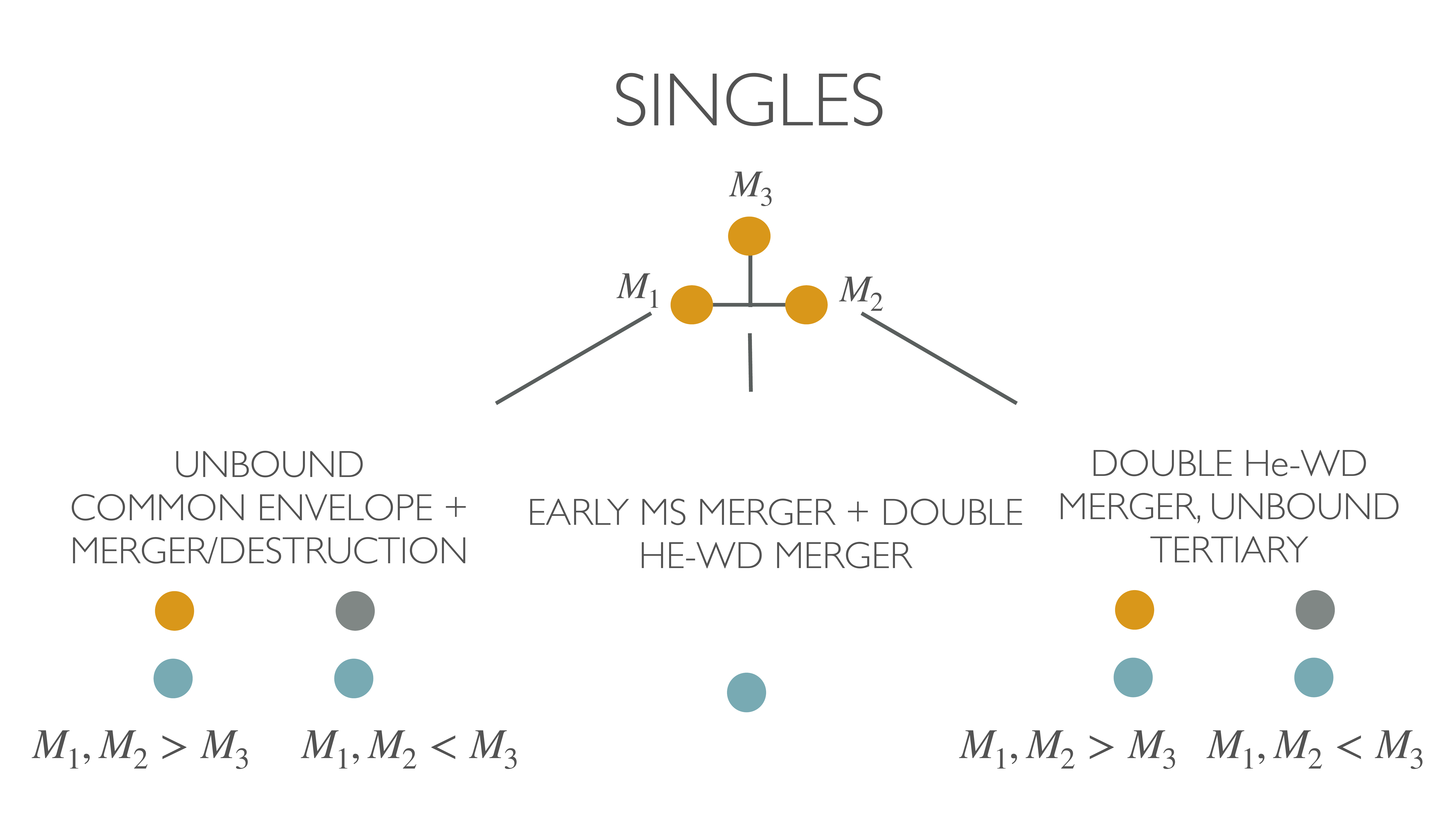}
 \caption{The formation channels identified for wide binaries.}
 \label{fig:sdbsinglesform}
\end{figure}

The wide binaries form when either the inner binary merges or one of the stars of the inner binary is destroyed. The outer tertiary remains bound to the system. An overview of the evolutionary channels to produce an sdOB star in a wide binary is show in Fig. \ref{fig:sdbwidebinariesform}. An overview of the evolutionary channels to produce a singleton sdOB star is shown in Fig. \ref{fig:sdbsinglesform}. Here wide binaries refer to companions sufficiently far away that they are non-interacting.

\subsubsection{He-WD + He-WD Mergers}
The inner binary consists of two stars with initial masses between $1\,\rm{M_\odot}$ and $2\,\rm{M_\odot}$. The initially more massive star evolves onto the RGB wherein mass-transfer takes place and removes the RGB envelope. Unlike with sdOB formation, He is not ignited in the core and the star instead becomes a He-WD. The secondary of the inner binary then evolves onto the RGB and also loses its envelope via RLOF. Typically two phases of common-envelope occur. The resulting close He-WD binary loses orbital energy via gravitational wave radiation. The two He-WDs merge within the Hubble time. He is ignited in the core of the merged product to create an sdOB. 

If the outer tertiary remains bound to the system a wide binary is formed. If the outer tertiary is unbound during the evolution of the system a singleton sdOB is born. The outer tertiary can be either a WD or MS star depending on its relative initial mass.

\subsubsection{Double Merger}
Two mergers can occur if the triple is initially very tight. The first merger occurs in the inner binary during the main-sequence evolution. The resulting close binary continues to evolve into a double He-WD binary via two phases of common-envelope on the RGB. The orbit of the double He-WD binary shrinks via gravitational wave emission until the two objects merge. A single sdOB is formed when He ignites in the core.

\subsubsection{Common Envelope + Merger}
This channel is similar to the previously described common-envelope channels. The more massive star of the inner binary evolves onto the RGB. The RGB overflows its Roche lobe and unstable mass-transfer begins. A common envelope is formed. During the common-envelope evolution the He core of the RGB star merges with its main-sequence companion. The envelope is blown off the post-common-envelope, post-merger object. He ignites in the core to form an sdOB. 

If the outer tertiary remains bound the sdOB is in a wide binary, if the outer tertiary is unbound the sdOB is a singleton. This formation channel is somewhat dubious. Recent 3D hydro-dynamical simulations carried out by \cite{2021MNRAS.500.1921G} suggest that the envelope of the merged product is unlikely to be fully expelled. 

\subsubsection{Destruction of a low-mass Main Sequence Companion}
This channel is another variant of common-envelope evolution in the inner binary. This channel only occurs for low mass-main sequence companions with initial masses below $0.7\,\rm{M_\odot}$ which are close to fully convective. The more massive star of the inner binary evolves onto the RGB and common-envelope ejection occurs. After the common-envelope evolution, the companion also experiences some Roche Lobe overflow, owing to the shrunken orbit. The main-sequence companion star loses its small outer radiative region. Once the convective core is exposed mass transfer proceeds on a dynamical time-scale. The low-mass star is assumed to be destroyed \citet{2002MNRAS.329..897H}. The matter from the low-mass star is ejected too rapidly to be accreted by the companion star. Once again, the tertiary can remain bound or be unbound to give a wide binary or single system.

\section{Statistical Outcomes}
We show the initial conditions of the systems which form sdOBs for both the binary and triples case. We further show the distributions of the formed sdOB systems, again for the binary and triple case. The role of ZLK oscillations in formation is investigated.
\subsection{Distributions of the sdOBs}
\begin{figure}
 \includegraphics[width=\columnwidth]{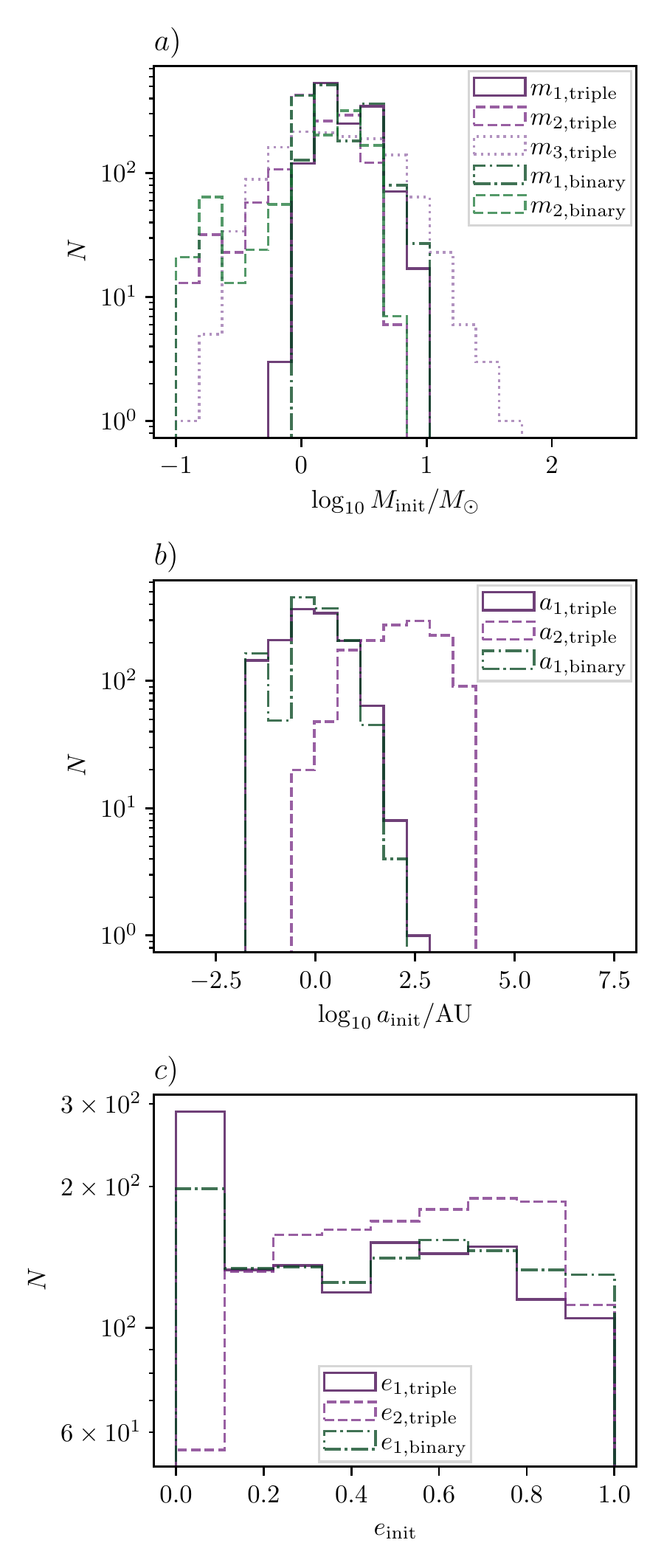}
 \caption{The initial conditions of the systems which produced an sdOB star. The purple lines refer to the triple case and the green lines refer to the binary case. Subplot $a)$ shows the masses of the stars where $m_1$ is the most massive star in the inner binary, $m_2$ is the least massive star in the inner binary and $m_3$ is the mass of the tertiary. Subplot $b)$ are the semi-major axes of the inner binary, $a_1$, and outer tertiary orbit, $a_2$. Subplot $c)$ are the eccentricities of the inner binary, $e_1$, and the outer tertiary, $e_2$. The same distributions for $m_1$, $m_2$, $a_1$ and $e_1$ are used for the data set with the outer tertiary removed.}
 \label{fig:sdbbinaries}
\end{figure}

\begin{figure}
 \includegraphics[width=\columnwidth]{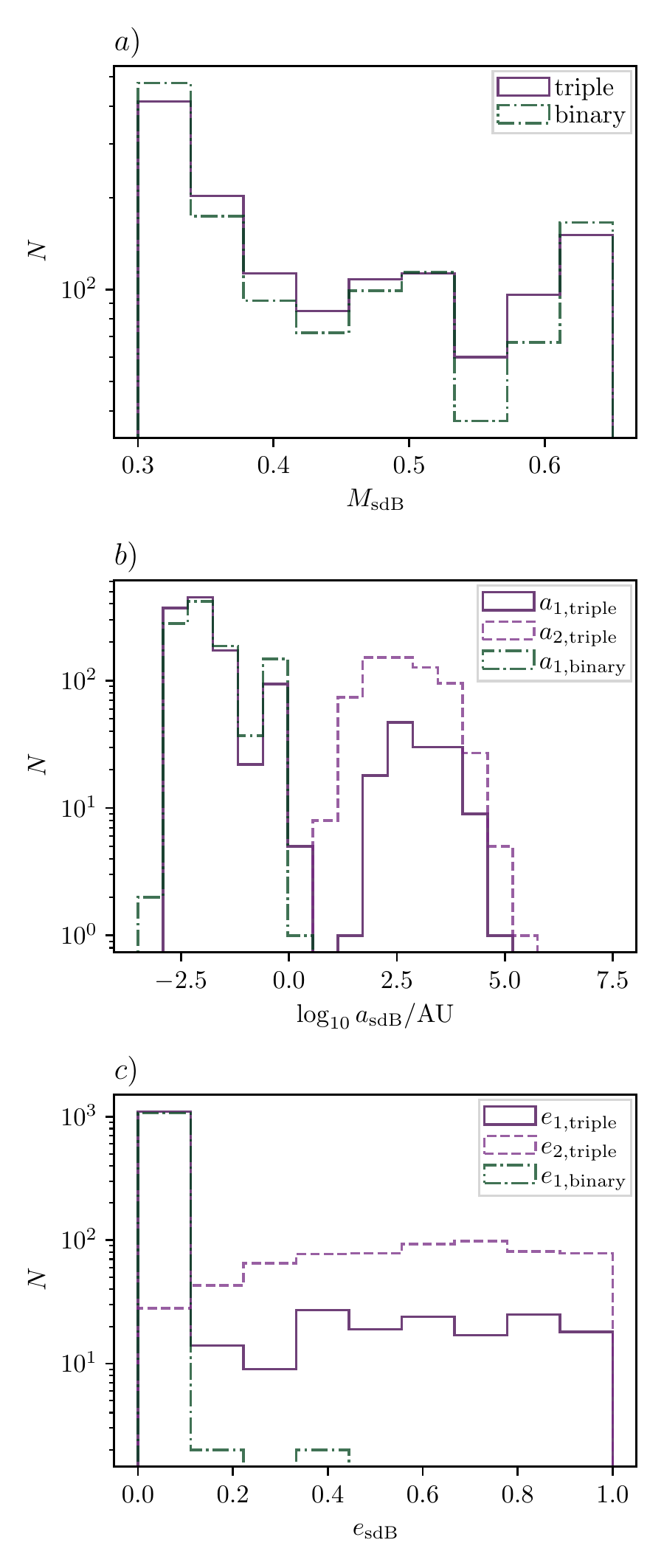}
 \caption{The distributions of properties of the formed sdOBs identified in this paper. The purple lines refer to the triple case and the green lines refer to the binary case. Subplot $a)$ shows the masses of the sdOBs. Subplot $b)$ are the semi-major axes of the inner binary, $a_1$, and outer tertiary orbit, $a_2$. Subplot $c)$ are the eccentricities of the inner binary, $e_1$, and the outer tertiary, $e_2$.}
 \label{fig:sdbtriples}
\end{figure}

Outcomes and initial parameters for the formed sdOBs are shown in Figs \ref{fig:sdbbinaries} (ICs) and \ref{fig:sdbtriples} (sdOB parameters).  Owing to the relatively small number of objects of interest formed, and the large number of channels identified current statistics are approximate. We consider all channels together and leave detailed analysis of individual channels for future work. When the inner binary merges but the outer tertiary stays bound to the system the semi-major axis of the binary is labelled by $a_1$. The sdOB binaries with $a_1 > 50 \, \rm{AU}$ are such systems. The inner binary of the sdOB containing systems have a slight preference for close orbits when comparing the triple and binary case. More initially circular systems lead to sdOBs in the triple case than the binary case because secular evolution can excite eccentricity and thus enhance sdOB formation. 

The mass distributions of the formed sdOBs are similar when comparing the binary and triple case. A small excess of higher mass objects can be seen in the sdOB mass in the triple case. Three peaks can be seen in the distribution. Theoretical results also find three different mass ranges for sdOBs. Stripped RGB stars which ignite He degenerately, and so have initial masses below $2.25\,M_\odot$, create sdOBs with masses of $0.47\,M_\odot$. Stripped RGB stars which ignite He non-degenerately are predicted to have masses $\sim 0.3\,M_\odot$ solar masses. Merger products have masses in the ranging from $0.3 \,M_\odot$ to $0.65 \,M_\odot$. The results of our simulations do not match these predictions. We find low-mass degenerate progenitors which produces sdOBs with masses $\sim 0.3\,M_\odot$ and non-degenerate progenitors which produce systems with canonical masses of $0.47\,M_\odot$. MSE uses the SSE results for handling the stellar evolution \citep{sse}. Comparison with results from BSE \citet{2002MNRAS.329..897H} reveals the same general result as MSE.  

The distribution of the semi-major axis of the inner binaries have three peaks. The peak centred around $\log_{10}a_1/\rm{AU} = -2 $ corresponds to the post common-envelope systems, the peak around $\log_{10}a_1/\rm{AU} = 0 $ to the stable Roche lobe overflow systems and the peak around $\log_{10}a_1/\rm{AU} = 3 $ to the systems where the inner binary has merged but the outer tertiary has remained bound to the merger product. These very wide binary systems also typically have $e>0$. The wide sdOBs channel is unique to triples, binary systems are unable to create sdOBs if the companion is non-interacting. The triple case has a preference for common-envelope evolution in comparison to the binary case. More stable RLOF systems are created in the binary case than the triple case.

Table \ref{tab:config} shows the fraction of systems in either a triple, binary or single at the time of sdOB formation. The binary case is more likely to lead to singleton sdOBs being formed. The triple case is equally likely to lead to sdOBS in binaries or triples. More triples are formed but they are often subject to dynamical instabilities and merge with their close companion within $5\,\rm{Myrs}$ and so are not included in these statistics.

\begin{table}
    \centering
    \begin{tabular*}{\columnwidth}{l|ccc}
        Configuration &Triple (Final) & Binary (Final) & Single (Final)\\ \hline
         Triple (Initial) & $0.48 \pm 0.08$& $0.45 \pm 0.08$ & $0.07 \pm 0.03$  \\
         Binary (Initial) & -& $0.83 \pm 0.09$ & $0.17 \pm0.05$
    \end{tabular*}
    \caption{The fraction of sdOBs in either triples, binaries or singles at the time of sdOB formation for both initial triples and initial binaries. The sdOBs which are unbound but at least one other member of the system has survived are counted as singletons. Poisson errors are included.}
    \label{tab:config}
\end{table}

\subsection{Von Zeipel-Lidov-Kozai Oscillations}
\begin{figure}
 \includegraphics[width=\columnwidth]{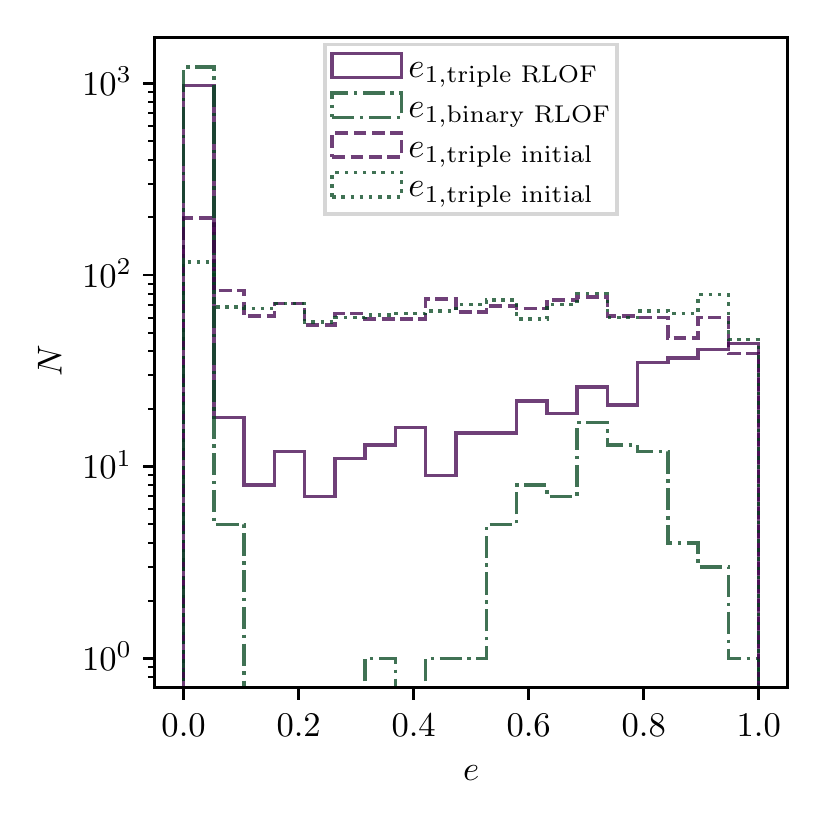}
 \caption{The distributions of the eccentricities for the systems which form sdOBs at the onset of the first phase of RLOF. The purple lines refer to the triple case and the green lines refer to the binary case. The initial conditions of the same systems are also shown for reference. }
 \label{fig:zlke}
\end{figure}

Examination of the periapsis distance in the evolution leading up to RLOF shows that approximately half of the hierarchical triples which form sdOBs undergo ZLK oscillations prior to the common-envelope evolution. In many cases, the ZLK oscillations excite the eccentricity of the inner orbit but do not reduce the periastron distance sufficiently to trigger RLOF. To investigate the role of ZLK oscillations on the formation of the sdOBs we look at the distribution of the eccentricity at the onset of the first phase of RLOF. Fig. \ref{fig:zlke} show the eccentricity distributions of the sdOB forming systems at the onset of the first episode of RLOF and the initial configuration for both the hierarchical triples and the inner binaries. When compared to the binary case, the triples show a clear increase of the number of systems with $e>0.05$. A small increase of the number of triple systems with $e_1>0.95$ can be seen when comparing to the initial conditions. Both points are taken as evidence that ZLK oscillations play a role in the formation of sdOBs in hierarchical triples. The binary systems with $e_1>0.4$ are the systems which form sdOBs via stable RLOF.

\subsection{Comparing Formation in Binary Case and Triple Case}
\begin{figure}
 \includegraphics[width=0.98\columnwidth]{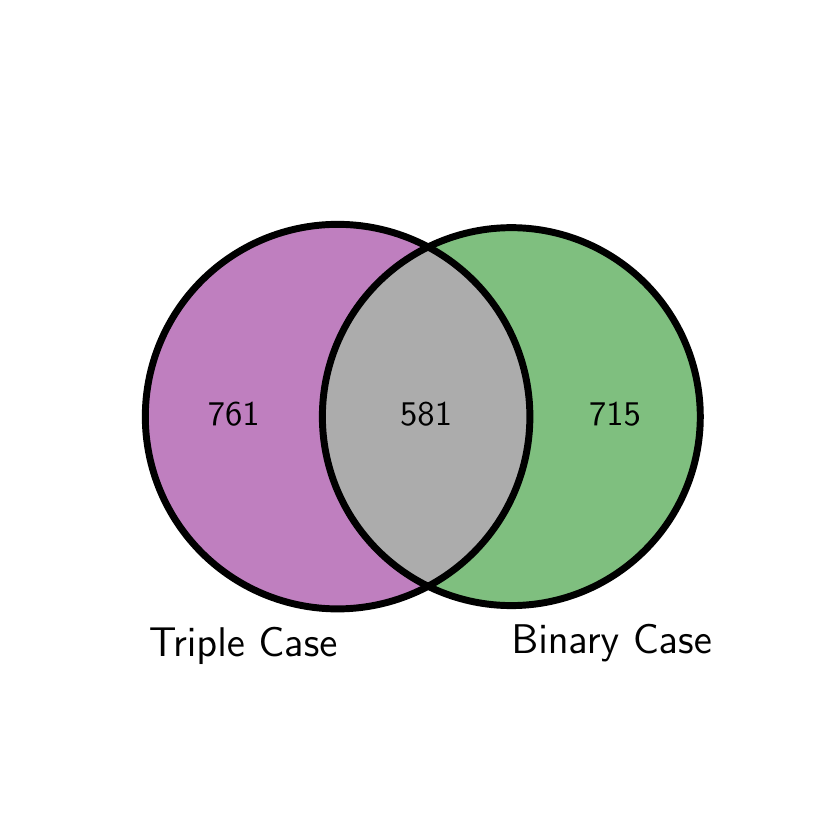}
 \caption{The number of systems where, for the same initial conditions, only the hierarchical triples create an sdOB (purple), only the inner binary create an sdOB (green) and both the hierarchical triple and inner binary create an sdOB (grey). }
 \label{fig:venn}
\end{figure}

To assess the role the outer tertiary plays in the above evolutionary channels we compare which systems form an sdOB in the binary case and the triple case. As shown in Fig. \ref{fig:venn}, only 43\% of the total sdOBs formed do so with or without the presence of an outer tertiary. The outer tertiary can either stabilise or destabilise the inner binary, thus deciding whether or not an sdOB is formed. The outer tertiary can initiate ZLK oscillations and excite the eccentricity of the inner binary. The increased eccentricity of the inner binary can trigger Roche Lobe overflow in a system that otherwise would not experience mass transfer. Alternatively, in systems where RLOF occurs, the eccentricity can become very large inducing common-envelope or a merger. The outer tertiary may stop the inner binary expanding and reaching such high eccentricity. Mass-transfer still occurs but not enough to remove the envelope of an RGB star. 

\section{Discussion}

We present the evolution channels found within our data set. Owing to the restricted initial mass distribution of the primary star in the inner binary we may have excluded some regions of parameter space which could produce sdOBs. Our statistics of the orbital parameters of the formed systems may be biased by our restricted initial conditions. 

About 8\% of the 100000 triple systems did not complete their evolution within the maximum allowed wall time of 10 hrs. All the inner binaries completed their evolution within the maximum wall time so the binary set is slightly larger than the triple set. We find 94 sdOBs are produced in the binary case where the corresponding model did not complete its evolution in the triple case.

Currently we only consider sdOBs produced from the merging of two He-WDs. Merger channels including CO-WDs have been proposed but are not considered in this work. The sdOBs from CO-WD mergers would likely be chemically peculiar. The CO-WD merger channel is not predicted to be a dominant channel for sdOB formation \citet{justhamco,marcellowd}. 

Some of sdOBs in hierarchical triple configurations should be observable, particularly if the companions are MS type and the outer tertiary is not in too wide an orbit. A close sdOB binary with a very distant, faint WD will likely not be observable as a triple. The inner binaries in the stable and unstable RLOF channels have different resulting semi-major axes. As shown in  Fig. \ref{fig:sdbtriples}, the CE systems have tighter orbits than the stable RLOF systems so they are expected to be observationally distinguishable from each other. The evolution mechanism may not be deducible from the system configuration of the sdOBs in close binaries as the envelope removal mechanism is similar in the systems with unbound tertiaries. Note that the mass transfer prescription used in population synthesis typically overestimates the number of common-envelope systems \citep{Chang_2022}.

\section{Conclusions}

In this work we present 14 distinct evolutionary pathways in which hierarchical triples can lead to sdOB formation. Each channel typically has several possible companion combinations. Twelve of the evolutionary pathways are binary evolution scenarios with an outer triple. In these cases either the envelope of a RGB star is removed or two He-WD stars merge. We additionally find two triple common envelope pathways. Both triple common envelope scenarios occur when the outer tertiary is more massive than both of the stars in the inner binary. The outer tertiary evolves more rapidly and on one of the giant branches forms an envelope which engulfs the inner binary. One triple common envelope evolution scenario proceeds as follows. A RGB star engulfs both of its companions, has its envelope removed then ignites He to form an sdOB star. Alternatively, a thermally pulsating AGB star can overflow its Roche Lobe and form a common envelope with the inner binary. The inner binary merges, evolves on to the RGB and goes through a second phase of common envelope evolution. The merged product has its envelope removed, ignites He in the core and becomes a sdOB star.  

We consider two model sets, one consisting of hierarchical triples, the second consisting of the same inner binary of the hierarchical triples but with the outer tertiary removed. The initial conditions of the inner binary of both model sets are identical. Comparison of the results from the inner binaries and the hierarchical triples show the influence of the outer tertiary. The overall number of formed sdOB stars is similar in both model sets. The presence of an outer tertiary alters which systems form sdOBs. In only 43\% of the systems considered do both the triple case and binary case form an sdOB star under the same initial conditions. The triple case creates more systems via common-envelope evolution and represses RLOF when compared to the binary case. The number of double He-WD mergers is not significantly affected by an outer tertiary. ZKL oscillations are present in around 50\% of the hierarchical triples which form sdOBs in the evolution prior to common-envelope. In some cases the eccentricity excitation from the ZKL oscillations triggers the mass-transfer which removes the envelope of an sdOB progenitor star.

The sdOBs in hierarchical triples are observable, particularly if the outer companion is still a MS star. If the outer tertiary is a WD star its presence will be more difficult to detect. We find outer tertiaries with semi-major axis distributed between 1 and $10^{5.5}\,\rm{AU}$. The sdOBs in wide binaries have semi-major axis distributed between 10 and $10^5\,\rm{AU}$. The outer tertiary components and very wide binary companions are typically in eccentric orbits. We expect more systems to be in triples than to have very wide binary components. 

The focus of this work was to identify possible evolutionary channels for hierarchical triples to produce sdOBs. We formed $\sim 1300$ sdOB stars in both the binary case and the triple case. Many of the identified channels only occurred a handful of times thus statistical inferences regarding the populations couldn't be drawn. We save detailed statistical analysis of individual channels and the model parameter dependence of our results for future work.

\section*{Acknowledgements}
A.S.H. thanks the Max Planck Society for 
support through a Max Planck Research Group.

\section*{Data Availability}
 	The data underlying this article will be shared on reasonable request to the corresponding author.



\bibliographystyle{mnras}
\bibliography{example} 





\bsp	
\label{lastpage}
\end{document}